# Y-Q and I-U Symmetries, and New Unified Lifetime Formulas of Hadrons

Yi-Fang Chang
Department of Physics, Yunnan University, Kunming, 650091, China
(e-mail: yifangchang1030@hotmail.com)

**Abstract**
Based on the Y-Q and I-U symmetries between mass and lifetime on the general SU(3) theory, we can derive the lifetime formulas of hyperons and mesons: $H = \tau = A[2U(U+1) - Q/2]$ and $\tau = A'[(1/2) + 2U(U+1) - Q/2 - Q^2/3]$, which agree better with experiments. It is a new method on lifetime of hadrons described by quantum numbers. They are symmetrical with the corresponding mass formulas, and can be unified for mass and lifetime. Further, these formulas may extend to describe masses and lifetime of heavy flavor hadrons.
**Key Words**: hadron, lifetime, symmetry, weak interaction, strong interaction, quantum number.
**PACS**: 11.30.-j; 12.10.-g; 12.70. +q; 14.20.Jn; 14.40.-n

**1. Symmetry between mass and lifetime**

The basic characteristics are mass and life for various particles. For the masses of hadrons there is a well-known GMO mass formula [1,2]:

$$M = M_0 + AY + B[I(I+1) - Y^2/4]. \qquad (1)$$

But, there must be M=m for baryons and M=$m^2$ for mesons. We corrected Eq.(1) to a little different formula [3-5]:

$$M = M_0 + AY + B[I(I+1) - Y^2/2]. \qquad (2)$$

Such not only baryons and mesons are unified M=m, and Eq.(2) agrees accurately with the experiments.

So far, the lifetimes of hadrons cannot usually derive the unified formula. Only some formulas are obtained for few particles, for example, the muon lifetime $\tau_\mu$ [6-8], in which a formula is [8]:

$$\frac{1}{\tau_\mu} = W_\mu = \frac{G^2 m_\mu^5}{192\pi^3}[1 - 8y + 8y^3 - y^4 - 12y^2 \ln y]. \qquad (3)$$

Here $y \equiv m_e^2/m_\mu^2$ is very small, and $G = (1.16637 \pm 0.00002) \times 10^{-11} MeV^{-2}$ is obtained. Moreover, Jacob and Sachs discussed mass and lifetime of unstable particles [9]. Bander, et al., proposed a mechanism, in which the reaction $D^0 \to$ s+d+gluon as a source for the difference in the lifetimes of the charged and neutral D mesons [10].

The SU(3) symmetry is very successful for hadrons composed of u, d and s quarks at low energy. In the known SU(3) theory, the first order broken operator is [6]:

$$H = H_{VSI} + H_{MSI} = O' + AU_3, \quad (4)$$

where $U_3 = Y - Q/2, O = O' + M(U^2, I^2, Y^2)$, so

$$H = O + AY - AQ/2 + BI(I+1) + CY^2 + DU(U+1). \quad (5)$$

The more general result is [11]:

$$H \approx H(3,1) = O + m_1 Y + m_2 I(I+1) + m_3 Y^2 - \mu_1 Q + \mu_2 U(U+1) + O'. \quad (6)$$

For the masses of hadrons, when the operators Q and U of electromagnetic interaction are neglected, let C=-B/4, the GMO mass formula (1) is derived. If let C=-B/2, we derived a new modified accurate mass formula (2). Further, Eq.(2) can be obtained based on the Lagrangian of symmetry and its dynamical breaking or Higgs breaking [5].

**2. Lifetime formulas of hyperons and mesons**

If we suppose that Eq.(5) or (6) may be applied to the lifetime, because the strong interaction and corresponding operators Y and I may be neglected in the usual decays of hadrons, let O=0 and D/2=A, then a simplest lifetime formula

$$H = \tau = A[2U(U+1) - Q/2] \quad (7)$$

is obtained. The formula (7) is analogue to the mass splitting formula of electromagnetic interaction. We have known that $\Xi^0$ and n are U-spin triplet U=1, $\Xi^-, \Sigma^-$ and $\Sigma^+$, p are all U-spin doublet U=1/2, $\Lambda$ and $\Sigma^0$ are the mixture state of $U_0^0 - U_1^0$, $\Lambda = -\frac{1}{2}(U_0^0 - \sqrt{3}U_1^0)$, $\Sigma^0 = -\frac{1}{2}(\sqrt{3}U_0^0 + U_1^0)$. When $U_0^0 = 0, U_1^0 = 1$, so $U(\Lambda) = \frac{\sqrt{3}}{2}, U(\Sigma^0) = -\frac{1}{2}$, then the formula (7) agrees approximately with the lifetime of all hyperons (table 1), except $\Sigma^0$ (it is already the electromagnetic decay). For the mesons, if we take $K^0, \pi^-, K^+$, the same lifetime formula (7) will agree very well with the life of $K_L^0, \pi^-, K^+$ (table 2). Here we suppose that both mixture states $\pi^0, \eta$ are all U=0. It is difference between hyperons and mesons, and $\pi^0, \eta$ are also the electromagnetic decay. Then Eq.(7) will agree with the life of $\pi^0, \eta$, too.

If the lifetime formula becomes a symmetrical form with the mass formula (2), it will be

$$\tau = \tau_0 + A'[2U(U+1) - Q/2] + B'Q^2. \quad (8)$$

Let $\tau_0 = A'/2$ and $B' = -A'/3$, Eq.(8) becomes

$$\tau = A'[(1/2) + 2U(U+1) - Q/2 - Q^2/3], \tag{9}$$

then the agreement is better (table 1). From the symmetrical formulas between mass and lifetime, we obtain a lifetime relation of hyperons:

$$2[\tau(\Xi^0) + \tau(\Sigma^+)](7.404) = \tau(\Lambda) + 3\tau(\Xi^-)(7.549). \tag{10}$$

It is completely analogous to the Gell-Mann mass relation of baryons:

$$2[m(N) + m(\Xi)] = m(\Sigma) + 3m(\Lambda). \tag{11}$$

For the mesons, Eq.(7) may also become the symmetrical form:

$$\tau = A'U(U+1) + B'Q + C'Q^2 = A[2U(U+1) - Q/2] - BQ(Q+1), \tag{12}$$

then it agrees exactly with the experiments (table 2). First the lifetime of hadrons is determined by the kinds of decay-interaction and by the types of hadrons, both determine the order of magnitude of the parameters.

The exact lifetime formula of hyperons (table 1) is

$$\tau = A'[(1/2) + 2U(U+1) - Q/2] - B'Q^2 + C'Y(1+2Q). \tag{13}$$

The corresponding exact mass formula of hyperons is [4]

$$M = M_0 + AY + BI(I+1) + CY^2 + Q[d_1(Y + \frac{Q}{2}) - d_0]. \tag{14}$$

$$M = M_0 + AY + B[I(I+1) - Y^2/2] + Q^2(c_1 + c_2Y). \tag{15}$$

The decay of $J^P = (3/2)^+$ decimet baryons (except $\Omega^-$) is strong interaction, the lifetime (width) formula is

$$\Gamma = \Gamma_0(2^{2I}I) = 2\Gamma_0[I(I+1) + I^2Y]. \tag{16}$$

This agrees very well (table 3). Because I=1+(Y/2), Eq.(16) may change to representation of Y, it is symmetrical with (2).

When we extend Eq.(5) to the second approximation

$$H(33,11) = O + a_1Y - a_2Q + b_1I(I+1) + b_2U(U+1) + c_1Y^2 + c_2YQ +$$
$$c_3YI(I+1) + c_4YU(U+1) + c_5Q^2 + c_6QI(I+1) + c_7QU(U+1) + \quad . \tag{17}$$
$$c_8[I(I+1)]^2 + c_9I(I+1)U(U+1) + c_{10}U^2(U+1)^2$$

Various terms of the preceding exact formulas are the second approximate terms.

The mass and lifetime formulas (2) and (7) are just taken terms of supplement each other in Eq.(5). They are determined by strong interaction and corresponding Y-I symmetry, and by weak (electromagnetic) interaction and corresponding Q-U symmetry, respectively. The both aspects are again a replacement-symmetry of $Q \leftrightarrow -Y$ (-Q is the corresponding hypercharge of U-spin) and $I \leftrightarrow U$. For the first approximation, these masses of hadrons whose Y and I are the same are approximate equal, which is a degeneracy of u and d in the quark model, and both quarks are Y=1/3 and I=1/2. Similar, these lifetime of hadrons whose Q and U are the same are approximate equal, which is a degeneracy of d and s in the quark model, and both quarks are Q=-1/3 and U=1/2.

Because three SU(2) subgroups (iso-spin, V-spin and U-spin groups) are symmetrical, mass and lifetime of hadrons are classified according to isospin and U-spin, and both are the scalars of isospin and U-spin groups in mathematical representation, respectively. Therefore, mass and lifetime are symmetrical, and are unified by SU(3) theory. But the values of Y and Q are reverse, so the signs of Y and Q are also reverse in Eqs.(5)(6)(17). In the mass formulas of hadrons and the lifetime formulas of hyperons-mesons, these coefficients of $Y, Y^2$ and $Q, Q^2$ are all negative, and these coefficients of I(I+1) and U(U+1) are all positive, both aspects are symmetrical. In the second approximate formula (17) and in the exact formulas (14)(13), mass and lifetime are symmetrical, too.

Table 1. Lifetime of $J^p = (1/2)^+$ Hyperons

|  | Calc.Value | | | | | Obs.Value[12] |
|---|---|---|---|---|---|---|
|  | U | (7) A=0.8 | (8) $2\tau_0 = A' = 0.68$, $B' = -0.22$ | (9) $A' = 0.69$ | (13) $A' = 0.7$, $B' = 0.255$, $C' = 0.17$ | ($10^{-10}$ sec) |
| $\Xi^0$ | 1 | 3.2 | 3.060 | 3.105 | 2.98 | $2.90 \pm 0.09$ |
| $\Lambda$ | $\sqrt{3}/2$ | 2.586 | 3.538 | 2.575 | 2.6124 | $2.632 \pm 0.020$ |
| $\Xi^-$ | 1/2 | 1.6 | 1.48 | 1.495 | 1.665 | $1.639 \pm 0.015$ |
| $\Sigma^-$ | 1/2 |  |  |  | 1.495 | $1.479 \pm 0.011$ |
| $\Sigma^+$ | 1/2 | 0.8 | 0.8 | 0.805 | 0.795 | $0.8018 \pm 0.0026$ |
| $\Sigma^0$ | -1/2 | -0.4 | 0 | 0 | 0 | 0 |

Table 2. Lifetime of $J^{pc} = 0^{-+}$ Mesons

|  | Calc.Value | | | Obs.Value[12] |
|---|---|---|---|---|
|  | U | (7) A=1.3 | (12) $A' = 1.3$, $B' = 0.03$ | ($10^{-8}$ sec) |
| $K_L^0$ | 1 | 5.2 | 5.2 | $5.17 \pm 0.04$ |
| $\pi^-$ | 1/2 | 2.6 | 2.6 | $2.6029 \pm 0.0023$ |
| $K^+$ | 1/2 | 1.3 | 1.24 | $1.2384 \pm 0.0024$ |
| $\pi^0$ | 0 | 0 | 0 | 0 |
| $\pi^0$ | 0 | 0 | 0 | 0 |

Table 3. Width (MeV) of $J^p = (3/2)^+$ Baryons

|  | I | Calc.Value (16) $\Gamma_0 = 9.6$ | Obs.Value (Average) |
|---|---|---|---|
| $\Delta$ | 3/2 | 115.2 | 115-125($120 \pm 5$) |
| Σ | 1 | 38.4 | (+)35.8 ± 0.8 |
| | | | (0)36 ± 5 (37.1 ± 2.3) |
| | | | (-)39.4 ± 2.1 |
| Ξ | 1/2 | 9.6 | (0)9.1 ± 0.5 |
| | | | (-)9.9 ± 1.9(9.5 ± 0.4) |
| $\Omega^-$ | 0 | 0 | 0 |

**3. Applications of formulas in heavy flavor hadrons**

According to the symmetry of s-c quarks, the heavy flavor hadrons which made of u,d and c quarks may be classified by SU(3) octet and decuplet, we derived that these masses of the octet should obey corresponding GMO mass formula [5]:

$$M = M_0 + AC + B[I(I+1) - \frac{C^2}{4}], \qquad (18)$$

or new accurate mass formula

$$M = M_0 + AC + B[I(I+1) - \frac{C^3}{2}]. \qquad (19)$$

Using these formulas we described masses of heavy flavor hadrons, and predict masses of some hadrons [5].

For the masses of resonances, there is a well-known Regge trajectory formula:

$$S = m^2 = AJ + B. \qquad (20)$$

According to symmetry and unification of mass and lifetime (width), we assume that the width formula of corresponding resonances is [4]:

$$\Gamma = aJ + b. \qquad (21)$$

In order to agree with the experiments, it is not $\Gamma^2$. The formulas (20) and (21) may be derived from the simple harmonic oscillator model.

Further, these lifetime formulas are extended to heavy flavor hadrons. Based on the harmonic oscillator model, the lifetime formulas should be:

$$\tau = \tau_0 + aC, \qquad (22)$$

$$\tau = \tau_0 + bB, \qquad (23)$$

and so on. According to the formula (13) and Table 1, the sequence of lifetime of hyperons is $\Xi^0 > \Lambda > \Xi^- > \Sigma^- > \Sigma^+ > \Sigma^0$. Corresponding sequence of lifetime of heavy flavor baryons should be $\Xi_c^+ > \Lambda_c^+ > \Xi_c^0 > \Sigma_c^0 > \Sigma_c^{++} > \Sigma_c^+$, which agrees with experimental data [12]. But, $\tau(\Xi_c^+, \Lambda_c^+, \Xi_c^0) \sim 10^{-13} s$ and $\tau(\Sigma_c^0, \Sigma_c^{++}, \Sigma_c^+) \sim 10^{-23} s$.

For $\Xi_c^+$ (U=1), $\Lambda_c^+$ (U=1) and $\Xi_c^0$ (U=1/2), we propose a lifetime formula:

$$\tau = \{1.5[U(U+1) + Q] - 2.5Y\} \times 10^{-13}. \qquad (24)$$

Then $\tau(\Lambda_c^+)$=2, $\tau(\Xi_c^+)$=4.5, $\tau(\Xi_c^0)$=1.125, and the experimental data are $(2.00\pm0.06)$, $(4.42\pm0.26)$, $(0.98\pm0.23)\times10^{-13}$. They agree within the range of error. This formula includes the lifetime of $\Sigma_c^0$, $\Sigma_c^{++}$ and $\Sigma_c^+$, it must become

$$\tau = \{A[U(U+1)+Q]-BY\}(1-I). \tag{25}$$

From these formulas some equal-ratio relations may be derived:

$$\frac{\tau(D^\pm = d\bar{c})}{\tau(\psi = c\bar{c})} = \frac{1.051\times10^{-12}}{0.757\times10^{-20}} = 1.3884\times10^8 \cong \frac{\tau(B^\pm = u\bar{b})}{\tau(Y = b\bar{b})} = \frac{1.674\times10^{-12}}{1.254\times10^{-20}}$$
$$= 1.335\times10^8 \approx \frac{\tau(K_S^0 = d\bar{s})}{\tau(\eta = s\bar{s}\,?!)} = \frac{8.93\times10^{-11}}{5.578\times10^{-19}} = 1.601\times10^8. \tag{26}$$

$$\frac{\tau(\Lambda = uds)}{\tau(\Sigma^0 = uds)} = \frac{2.632\times10^{-10}}{7.4\times10^{-20}} = 3.557\times10^9 \cong$$
$$\frac{\tau(\Lambda_c^+ = udc)}{\tau(\Sigma_c^0 = ddc)} = \frac{2.0\times10^{-13}}{5.863\times10^{-23}} = 3.411\times10^9. \tag{27}$$

From this we may predict

$$\frac{\tau(D_t^\pm = d\bar{t})}{\tau(\psi = t\bar{t})} \cong 1.3\times10^8 \quad \text{and} \quad \frac{\tau(\Lambda_b^0 = udb)}{\tau(\Sigma_b)} \approx 3.48\times10^9. \tag{28}$$

In some hyperons and mesons, there are equal-ratio relations:

$$\frac{\tau(\Xi^0 = uss)}{\tau(\Sigma^- = dds)} = 1.959 \approx \frac{\tau(\Xi^- = dss)}{\tau(\Sigma^+ = uus)} = 2.044 \approx$$
$$\frac{\tau(K_L^0 = d\bar{s})}{\tau(\pi^\pm = d\bar{u})} = 1.986 \approx \frac{\tau(\pi^\pm)}{\tau(K^\pm = u\bar{s})} = 2.102. \tag{29}$$

For $J^P = (3/2)^+$ decimet baryons, the ratio relations are

$$\frac{\Gamma(\Delta)}{\Gamma(\Sigma)} = 3, \frac{\Gamma(\Sigma)}{\Gamma(\Xi)} = 4. \tag{30}$$

For decimet of heavy flavor hadrons, a corresponding width formula should be

$$\Gamma = \Gamma_0[I(I+1)+I^2C]. \tag{31}$$

In a word, we propose the lifetime formulas, which is a new method on lifetime of hadrons represented by quantum numbers. And some basic properties (mass and lifetime) of hadrons may be unified by the symmetry theory.

**References**

1.M.Gell-Mann, Phys.Rev. 125,1067(1962).
2.S.Okubo, Progr.Theoret.Phys. 27,949(1962).

3.Yi-Fang Chang, Hadronic J. 7,1118(1984).

4.Yi-Fang Chang, New Research of Particle Physics and Relativity. Yunnan Science and Technology Press. 1989. Phys.Abst. 93,No.1371(1990).

5.Yi-Fang Chang, Mass Formulas Derived by Symmetry Breaking and Prediction of Masses on Heavy Flavor Hadrons. arXiv:0803.0087.

6.B.T.Feld, Models of Elementary Particle. Blaisdell Publishing Company. 1969.

7.R.E.Marshak, Riazuddin and C.P.Ryan, Theory of Weak Interactions in Particle Physics. Wiley-Interscience. 1969.

8.W.Greiner and B.Muller, Gauge Theory of Weak Interactions. 3 Edition. Springer-Verlag. 2003.

9.R.Jacob and R.G.Sachs, Phys.Rev. 121,350(1961).

10.M.Bander, D.Silverman and A.Soni, Phys.Rev.Lett. 44,7(1980).

11.G.Zweig, Symmetries in Elementary Particle Physics. Academic. New York. 1964.

12.W-M Yao, C.Amsler, D.Asner, et al. Particle Data Group. J.Phys. G33,1(2006).